\documentclass[journal,final]{IEEEtran}
\usepackage{amsmath,amssymb,url,amsfonts,booktabs,multirow,cite,subfigure,graphicx,bm}
\usepackage[lined,ruled,commentsnumbered]{algorithm2e}
\allowdisplaybreaks \newtheorem{Definition}{Definition}
\DeclareMathOperator*{\argmin}{\arg\!\min}
\DeclareMathOperator*{\argmax}{\arg\!\max}
\newcommand{\tabincell}[2]{\begin{tabular}{@{}#1@{}}#2\end{tabular}}

\begin{document}
\title{Transfer Learning for EEG-Based Brain-Computer Interfaces: A Review of Progress Made Since 2016}
\author{
Dongrui~Wu, Yifan~Xu and Bao-Liang~Lu
\thanks{
D.~Wu and Y. Xu are with the Ministry of Education Key Laboratory of Image Processing and Intelligent Control, School of Artificial Intelligence and Automation, Huazhong University of Science and Technology, Wuhan 430074, China. Email: drwu@hust.edu.cn, yfxu@hust.edu.cn.    }
\thanks{
B.-L.~Lu is with the Center for Brain-Like Computing and Machine Intelligence, Department of Computer Science and Engineering, Key Laboratory of Shanghai Education Commission for Intelligent Interaction and Cognitive Engineering, Brain Science and Technology Research Center, and Qing Yuan Research Institute, Shanghai Jiao Tong University, 800 Dong Chuan Road, Shanghai 200240, China. Email: bllu@sjtu.edu.cn.    }
}

\maketitle

\begin{abstract}
A brain-computer interface (BCI) enables a user to communicate with a computer directly using brain signals.
The most common non-invasive BCI modality, electroencephalogram (EEG), is sensitive to noise/artifact and suffers between-subject/within-subject non-stationarity. Therefore, it is difficult to build a generic pattern recognition model in an EEG-based BCI system that is optimal for different subjects, during different sessions, for different devices and tasks. Usually, a calibration session is needed to collect some training data for a new subject, which is time-consuming and user unfriendly. Transfer learning (TL), which utilizes data or knowledge from similar or relevant subjects/sessions/devices/tasks to facilitate learning for a new subject/session/device/task, is frequently used to reduce the amount of calibration effort. This paper reviews journal publications on TL approaches in EEG-based BCIs in the last few years, i.e., since 2016. Six paradigms and applications -- motor imagery, event-related potentials, steady-state visual evoked potentials, affective BCIs, regression problems, and adversarial attacks -- are considered. For each paradigm/application, we group the TL approaches into cross-subject/session, cross-device, and cross-task settings and review them separately. Observations and conclusions are made at the end of the paper, which may point to future research directions.
\end{abstract}

\section{Introduction}

A brain-computer interface (BCI) enables a user to communicate with a computer using his/her brain signals directly \cite{Wolpaw2002,Lance2012}. The term was first coined by Vidal in 1973 \cite{Vidal1973}, although it had been studied previously \cite{Fetz1969,Delgado1969}. BCIs were initially proposed for disabled people \cite{Pfurtscheller2008}, but their current application scope has been extended to able-bodied users \cite{Erp2012}, in gaming \cite{Marshall2013}, emotion recognition \cite{Zheng2015}, mental fatigue evaluation \cite{Monteiro2019}, vigilance estimation \cite{Zheng2017a,Shi2013}, etc.

There are generally three types of BCIs \cite{Rao2013}:
\begin{enumerate}
\item \emph{Non-invasive BCIs}, which use non-invasive brain signals measured outside of the brain, e.g., electroencephalograms (EEGs) and functional near-infrared spectroscopy (fNIRS).
\item \emph{Invasive BCIs}, which require surgery to implant sensor arrays or electrodes within the grey matter under the scalp to measure and decode brain signals (usually spikes and local field potentials).
\item \emph{Partially invasive (semi-invasive) BCIs}, in which the sensors are surgically implanted inside the skull but outside the brain rather than within the grey matter.
\end{enumerate}

This paper focuses on non-invasive BCIs, particularly EEG-based BCIs, which are the most popular type of BCIs due to their safety, low cost, and convenience.

A closed-loop EEG-based BCI system, shown in Fig.~\ref{fig:BCI}, consists of the following components:

\begin{figure}[htpb] \centering
\includegraphics[width=\linewidth,clip]{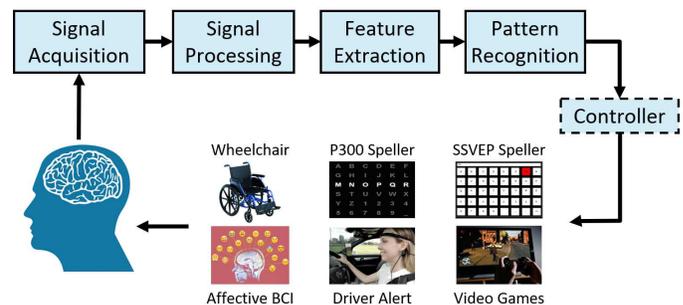}
\caption{Flowchart of a closed-loop EEG-based BCI system.} \label{fig:BCI}
\end{figure}

\begin{enumerate}
\item \emph{Signal acquisition} \cite{Liao2012}, which uses an EEG device to collect EEG signals from the scalp. In the early days, EEG devices used wired connections and gel to increase  conductivity. Currently, wireless connections and dry electrodes are becoming increasingly popular.
\item \emph{Signal processing} \cite{Makeig2012}, which usually includes temporal filtering and spatial filtering. The former typically uses a bandpass filter to reduce interference and noise, such as muscle artefacts, eye blinks, and DC drift. The latter combines different EEG channels to increase the signal-to-noise ratio. Popular spatial filters include common spatial patterns (CSP) \cite{Ramoser2000}, independent component analysis (ICA) \cite{Makeig1996a}, blind source separation \cite{Jung2000}, xDAWN \cite{Rivet2009}, etc.
\item \emph{Feature extraction}, for which time domain, frequency domain \cite{Wang2014b}, time-frequency domain, Riemannian space \cite{Yger2017} and/or functional brain connectivity \cite{Wu2019} features could be used.
\item \emph{Pattern recognition}. Depending on the application, a classifier or regression model is used.
\item \emph{Controller}, which outputs a command to control an external device, e.g., a wheelchair or a drone, or to alter the behaviour of an environment, e.g., the difficulty level of a video game. A controller may not be needed in certain applications, e.g., BCI spellers.
\end{enumerate}
When deep learning is used, feature extraction and pattern recognition can be integrated into a single neural network, and both components are optimized simultaneously and automatically.

EEG signals are weak, easily contaminated by interference and noise, non-stationary for the same subject, and varying across different subjects and sessions. Therefore, it is challenging to build a universal machine learning model in an EEG-based BCI system that is optimal for different subjects, during different sessions, for different devices and tasks. Usually, a calibration session is needed to collect some training data for a new subject, which is time-consuming and user unfriendly. Therefore, reducing this subject-specific calibration is critical to the market success of EEG-based BCIs.

Different machine learning techniques, e.g., transfer learning (TL) \cite{Pan2010} and active learning \cite{Settles2009}, have been used for this purpose. Among them, TL is particularly promising because it can utilize data or knowledge from similar or relevant subjects/sessions/devices/tasks to facilitate learning for a new subject/session/device/task. Moreover, it may also be integrated with other machine learning techniques, e.g., active learning \cite{drwuPLOS2013,drwuTNSRE2016}, for even better performance. This paper focuses on TL in EEG-based BCIs.

There are three classic classification paradigms in EEG-based BCIs, which will be considered in this paper:
\begin{enumerate}
\item \emph{Motor imagery (MI)} \cite{Pfurtscheller2001}, which can modify neuronal activity in primary sensorimotor areas, is similar to a real executed movement. As different MIs affect different regions of the brain, e.g., the left (right) hemisphere for right-hand (left-hand) MI  and centre for feet MI, a BCI can decode MI from the EEG signals and map it to a specific command.
\item \emph{Event-related potentials (ERP)} \cite{Handy2005,Lees2018}, which are any stereotyped EEG responses to a visual, audio, or tactile stimulus. The most frequently used ERP is P300 \cite{Sutton1965}, which occurs approximately 300 ms after a rare stimulus.
\item \emph{Steady-state visual evoked potentials (SSVEP)} \cite{Friman2007}. The EEG oscillates at the same (or multiples of) frequency of the visual stimulus at a specific frequency, usually between 3.5 and 75 Hz \cite{Beverina2003}. This paradigm is frequently used in BCI spellers \cite{Chen2015a}, as it can achieve a very high information transfer rate.
\end{enumerate}

EEG-based affective BCIs (aBCIs) \cite{Muhl2014,AlNafjan2017,Shen2019,Alarcao2019}, which detect affective states (moods, emotions) from EEGs and use them in BCIs, have become an emerging research area. There are also some interesting regression problems in EEG-based BCIs, e.g., driver drowsiness estimation \cite{drwuFWET2019,drwuTITS2020,drwuTFS2017} and user reaction time estimation \cite{drwuRG2017}.
Additionally, recent research \cite{drwuBCIAttack2019,drwuTAR2019} has shown that BCIs also suffer from adversarial attacks, where deliberately designed tiny perturbations are added to benign EEG trials to fool the machine learning model and cause dramatic performance degradation. This paper also considers aBCIs, regression problems, and adversarial attacks of EEG-based BCIs.

Although TL has been applied in all of the above EEG-based BCI paradigms and applications, to our knowledge, there is no comprehensive and up-to-date review on it. Wang \emph{et al.} \cite{Wang2015} performed a short review in a conference paper in 2015. Jayaram \emph{et al.} \cite{Jayaram2016} gave a brief review in 2016, considering only cross-subject and cross-session transfers. Lotte \emph{et al.} \cite{Lotte2018} provided a comprehensive review of classification algorithms for EEG-based BCIs between 2007 and 2017. Again, they only considered cross-subject and cross-session transfers. Azab \emph{et al.} \cite{Azab2018} performed a review of four categories of TL approaches in BCIs in 2018: 1) instance-based TL; 2) feature-representation TL; 3) classifier-based TL; and 4) relational-based TL.

However, all the aforementioned reviews considered only cross-subject and cross-session TL of the three classic paradigms of EEG-based BCIs (MI, ERP and SSVEP) but did not mention the more challenging cross-device and cross-task transfers, aBCIs, regression problems and adversarial attacks.

To fill these gaps and to avoid overlapping too much with previous reviews, this paper reviews journal publications of TL approaches in EEG-based BCIs in the last few years, i.e., since 2016. Six paradigms and applications are considered: MI, ERP, SSVEP, aBCI, regression problems, and adversarial attacks. For each paradigm/application, we group the TL approaches into cross-subject/session (because these two concepts are essentially the same), cross-device, and cross-task settings and review them separately, unless no TL approaches have been proposed for that category. Some TL approaches may cover more than two categories, e.g., both cross-subject and cross-device transfers were considered. In this case, we introduce them in the more challenging category, e.g., cross-device TL. When there are multiple TL approaches in each category, we generally introduce them according to the years in which they were proposed, unless there are intrinsic connections among several approaches.

The remainder of this paper is organized as follows: Section~\ref{sect:TL} briefly introduces some basic concepts of TL. Sections~\ref{sect:MI}-\ref{sect:Attack} review TL approaches in MI, ERP, SSVEP, aBCIs, regression problems, and adversarial attacks, respectively. Section~\ref{sect:Conclusions} makes observations and conclusions, which may point to some future research directions.

\section{TL Concepts and Scenarios} \label{sect:TL}

This section introduces the basic definitions of TL, some related concepts, e.g., domain adaptation and covariate shift, and different TL scenarios in EEG-based BCIs.

In machine learning, a feature vector is usually denoted by a bold symbol $\bm{x}$. To emphasize that each EEG trial is a 2D matrix, this paper denotes a trial by $X\in\mathbb{R}^{E\times T}$, where $E$ is the number of electrodes and $T$ is the number of time domain samples. Of course, $X$ can also be converted into a feature vector $\bm{x}$.

\subsection{TL Concepts}

\begin{Definition} A \emph{domain} \cite{Pan2010,Long2014} $\mathcal{D}$ consists of a feature space $\mathcal{X}$ and its associated marginal probability distribution $P(X)$, i.e., $\mathcal{D}=\{\mathcal{X},P(X)\}$, where $X\in \mathcal{X}$.
\end{Definition}

A source domain $\mathcal{D}_s$ and a target domain $\mathcal{D}_t$ are different if they have different feature spaces, i.e., $\mathcal{X}_s\neq \mathcal{X}_t$, and/or different marginal probability distributions, i.e., $P_s(X)\neq P_t(X)$.

\begin{Definition} Given a domain $\mathcal{D}$, a \emph{task}\cite{Pan2010,Long2014} $\mathcal{T}$ consists of a label space $\mathcal{Y}$ and a prediction function $f(X)$, i.e., $\mathcal{T}=\{\mathcal{Y},f(X)\}$.
\end{Definition}

Let $y\in \mathcal{Y}$. Then, $f(X)=P(y|X)$ is the conditional probability distribution. Two tasks $\mathcal{T}_s$ and $\mathcal{T}_t$ are different if they have different label spaces, i.e., $\mathcal{Y}_s\neq \mathcal{Y}_t$, and/or different conditional probability distributions, i.e., $P_s(y|X)\neq P_t(y|X)$.

\begin{Definition} Given a source domain $\mathcal{D}_s=\{(X_s^i,y_s^i)\}_{i=1}^N$ and a target domain $\mathcal{D}_t$ with $N_l$ labelled samples $\{(X_t^i,y_t^i)\}_{i=1}^{N_l}$ and $N_u$ unlabelled samples $\{X_t^i\}_{i=N_l+1}^{N_l+N_u}$, \emph{transfer learning} (TL) aims to learn a target prediction function $f: X_t \mapsto y_t$ with low expected error on $\mathcal{D}_t$ under the general assumptions that $\mathcal{X}_s\neq \mathcal{X}_t$, $\mathcal{Y}_s\neq \mathcal{Y}_t$, $P_s(X)\neq P_t(X)$, and/or $P_s(y|X)\neq P_t(y|X)$.
\end{Definition}

In \emph{inductive TL}, the target domain has some labelled samples, i.e., $N_l>0$. For most inductive TL scenarios in BCIs, the source domain samples are labelled, but they could also be unlabelled. When the source domain samples are labelled, inductive TL is similar to \emph{multi-task learning} \cite{Zhang2018b}. The difference is that multi-task learning tries to learn a model for every domain simultaneously, whereas inductive TL focuses only on the target domain. In \emph{transductive TL}, the source domain samples are all labelled, but the target domain samples are all unlabelled, i.e., $N_l=0$. In \emph{unsupervised TL}, no samples in either domain are labelled.

Domain adaptation is a special case of TL, or more specifically, transductive TL:

\begin{Definition} Given a source domain $\mathcal{D}_s$ and a target domain $\mathcal{D}_t$, \emph{domain adaptation} aims to learn a target prediction function $f: \bm{x}_t \mapsto y_t$ with low expected error on $\mathcal{D}_t$, under the assumptions that $\mathcal{X}_s=\mathcal{X}_t$ and $\mathcal{Y}_s=\mathcal{Y}_t$, but $P_s(X)\neq P_t(X)$ and/or $P_s(y|X)\neq P_t(y|X)$.
\end{Definition}

Covariate shift is a special and simpler case of domain adaptation:
\begin{Definition} Given a source domain $\mathcal{D}_s$ and a target domain $\mathcal{D}_t$, \emph{covariate shift} occurs when $\mathcal{X}_s=\mathcal{X}_t$, $\mathcal{Y}_s=\mathcal{Y}_t$, $P_s(y|X)= P_t(y|X)$, but $P_s(X)\neq P_t(X)$.
\end{Definition}

\subsection{TL Scenarios}

According to the variations between the source and the target domains, there can be different TL scenarios in EEG-based BCIs:
\begin{enumerate}
{\item \emph{Cross-subject TL}.} Data from other subjects (the source domains) are used to facilitate the calibration for a new subject (the target domain). Usually, the task and EEG device are the same across subjects.
\item \emph{Cross-session TL}. Data from previous sessions (the source domains) are used to facilitate the calibration of a new session (the target domain). For example, data from previous days are used in the current calibration. Usually, the subject, task and EEG device are the same across sessions.
\item \emph{Cross-device TL}. Data from one EEG device (the source domain), is used to facilitate the calibration of a new device (the target domain). Usually, the task and subject are the same across EEG devices.
\item \emph{Cross-task TL}. Labelled data from other similar or relevant tasks (the source domains) is used to facilitate the calibration for a new task (the target domain). For example, data from left- and right-hand MI are used in the calibration of feet and tongue MI. Usually, the subject and EEG device are the same across tasks.
\end{enumerate}
Since cross-subject TL and cross-session TL are essentially the same, this paper combines them into one category: cross-subject/session TL. Generally, cross-device TL and cross-task TL are more challenging than cross-subject/session TL; hence, they were less studied in the literature.

The above simple TL scenarios could also be mixed to form more complex TL scenarios, e.g., cross-subject and cross-device TL \cite{drwuTNSRE2016}, cross-subject and cross-task TL \cite{drwuLA2020}, etc.



\section{TL in MI-Based BCIs} \label{sect:MI}

This section reviews recent progress in TL for MI-based BCIs. Many of them used the BCI Competition datasets\footnote{http://www.bbci.de/competition/}.

Assume there are $S$ source domains, and the $s$-th source domain has $N_s$ EEG trials. The $n$-th trial of the $s$-th source domain is denoted by $X_s^n\in \mathbb{R}^{E\times T}$, where $E$ is the number of electrodes and $T$ is the number of time domain samples from each channel. The corresponding covariance matrix is $C_s^n\in\mathbb{R}^{E\times E}$, which is symmetric and positive definite (SPD) and lies on a Riemannian manifold. For binary classification, the label for $X_s^n$ is $y_s^n\in\{-1,1\}$. The $n$-th EEG trial in the target domain is denoted by $X_t^n$, and the covariance matrix is denoted by $C_t^n$. These notations are used throughout the paper.

\subsection{Cross-Subject/Session Transfer} \label{sect:CSSMI}

Dai \emph{et al.} \cite{Dai2018} proposed the transfer kernel common spatial patterns (TKCSP) method, which integrates kernel common spatial patterns (KCSP) \cite{Albalawi2012} and transfer kernel learning (TKL) \cite{Long2015} for EEG trial spatial filtering in cross-subject MI classification. It first computes a domain-invariant kernel by TKL and then uses it in the KCSP approach, which further finds the components with the largest energy difference between two classes. Note that TL was used in EEG signal processing (spatial filtering) instead of classification.

Jayaram \emph{et al.} \cite{Jayaram2016} proposed a multi-task learning framework for cross-subject/session transfers, which does not need any labelled data in the target domain. The linear decision rule is $y=\mathrm{sign}(\boldsymbol{\mu}_{\boldsymbol{\alpha}}^\mathrm{T} X\boldsymbol{\mu}_{\bm{w}})$, where $\boldsymbol{\mu}_{\boldsymbol{\alpha}}\in \mathbb{R}^{C\times 1}$ are the channel weights and $\boldsymbol{\mu}_{\bm{w}}\in\mathbb{R}^{T\times 1}$ are the feature weights. $\boldsymbol{\mu}_{\boldsymbol{\alpha}}$ and $\boldsymbol{\mu}_{\bm{w}}$ are obtained by minimizing
\begin{align}
\min_{\boldsymbol{\alpha}_s,\bm{w}_s}
&\left[\frac{1}{\lambda}\sum_{s=1}^S\sum_{n=1}^{N_s}\left\|\boldsymbol{\alpha}_s^\mathrm{T} X_s^n\bm{w}_s-y_s^n\right\|^2 \right. \nonumber\\
&\left.+\sum_{s=1}^S \Omega(\bm{w}_s;\boldsymbol{\mu}_{\bm{w}},\boldsymbol{\Sigma}_{\bm{w}})
+\sum_{s=1}^S \Omega(\boldsymbol{\alpha}_s;\boldsymbol{\mu}_{\boldsymbol{\alpha}},
\boldsymbol{\Sigma}_{\boldsymbol{\alpha}})\right], \label{eq:Jayaram}
\end{align}
where $\boldsymbol{\alpha}_s\in \mathbb{R}^{C\times 1}$ are the channel weights for the $s$-th source subject, $\bm{w}_s\in\mathbb{R}^{T\times 1}$ are the feature weights, $\lambda$ is a hyperparameter, and $\Omega(\bm{w}_s;\boldsymbol{\mu}_{\bm{w}},\boldsymbol{\Sigma}_{\bm{w}})$ is the negative log prior probability of $\bm{w}_s$ given the Gaussian distribution parameterized
by $(\boldsymbol{\mu}_{\bm{w}},\boldsymbol{\Sigma}_{\bm{w}})$. $\boldsymbol{\mu}_{\bm{w}}$ and $\boldsymbol{\Sigma}_{\bm{w}}$ ($\boldsymbol{\mu}_{\boldsymbol{\alpha}}$ and $\boldsymbol{\Sigma}_{\boldsymbol{\alpha}}$) are the mean vector and covariance matrix of $\{\bm{w}_s\}_{s=1}^S$ ($\{\boldsymbol{\alpha_s}\}_{s=1}^S$), respectively.

In (\ref{eq:Jayaram}), the first term requires $\boldsymbol{\alpha}_s$ and $\bm{w}_s$ to work well for the $s$-th source subject; the second term ensures that the divergence of $\bm{w}_s$ from the shared $(\boldsymbol{\mu}_{\bm{w}},\boldsymbol{\Sigma}_{\bm{w}})$ is small, i.e., all the source subjects should have similar $\bm{w}_s$ values; and the third term ensures that the divergence of $\boldsymbol{\alpha}_s$ from the shared $(\boldsymbol{\mu}_{\boldsymbol{\alpha}}, \boldsymbol{\Sigma}_{\boldsymbol{\alpha}})$ is small. $\boldsymbol{\mu}_{\boldsymbol{\alpha}}$ and $\boldsymbol{\mu}_{\bm{w}}$ can be viewed as the subject-invariant characteristics of stimulus prediction and hence used directly by a new subject. Jayaram \emph{et al.} demonstrated that their approach worked well on cross-subject transfers in MI classification and cross-session transfers for one patient with amyotrophic lateral sclerosis.

Azab \emph{et al.} \cite{Azab2019} proposed weighted TL for cross-subject transfers in MI classification as an improvement of the above approach. They assumed that each source subject has plenty of labelled samples, whereas the target subject has only a few labelled samples. They first trained a logistic regression classifier for each source subject by using a cross-entropy loss function with an L2 regularization term. Then, a logistic regression classifier for the target subject was trained so that the cross-entropy loss of the few labelled samples in the target domain is minimized, and its parameters are close to those of the source subjects. The mean vector and covariance matrix of the classifier parameters in the source domains were computed in a similar way to that in \cite{Jayaram2016}, except that for each source domain, a weight determined by the Kullback-Leibler divergence between it and the target domain was used.

Hossain \emph{et al.} \cite{Hossain2018} proposed an ensemble learning approach for cross-subject transfers in multi-class MI classification. Four base classifiers were used, all constructed using TL and active learning: 1) multi-class direct transfer with active learning (mDTAL), a multi-class extension of the active TL approach proposed in \cite{drwuSMC2014}; 2) multi-class aligned instance transfer with active learning, which is similar to mDTAL except that only the source domain samples correctly classified by the corresponding classifier are transferred; 3) most informative and aligned instances transfer with active learning, which transfers only the source domain samples correctly classified by its classifiers and near the decision boundary (i.e., the most informative samples); and 4) most informative instances transfer with active learning, which transfers only source domain samples close to the decision boundary. The four base learners were finally stacked to achieve a more robust performance.

Since the covariance matrices of EEG trials are SPD and lie on a Riemannian manifold instead of in Euclidean space, Riemannian approaches \cite{Yger2017} have become popular in EEG-based BCIs. Different TL approaches have also been proposed recently.

Zanini \emph{et al.} \cite{Zanini2018} proposed a Riemannian alignment (RA) approach to centre the EEG covariance matrices $\{C_k^n\}_{n=1}^{N_k}$ in the $k$-th domain with respect to a reference covariance matrix $\overline{R}_k$ specific to that domain. More specifically, RA computes first the covariance matrices of some \emph{resting} trials in the $k$-th domain, in which the subject is not performing any task, and then calculates their Riemannian mean $\overline{R}_k$. $\overline{R}_k$ is next used as the reference matrix to reduce the inter-subject/session variation:
\begin{align}
  \widetilde{C}_k^n=\overline{R}_k^{-\frac{1}{2}}C_k^n\overline{R}_k^{-\frac{1}{2}}, \label{eq:RA}
\end{align}
where $\widetilde{C}_k^n$ is the aligned covariance matrix for $C_k^n$. Equation (\ref{eq:RA}) centres the reference state of different subjects/sessions at the identity matrix. In MI, the resting state is the time window during which the subject is not performing any task, e.g., the transition window between two MI tasks. In ERP, the non-target stimuli are used as the resting state, requiring that some labelled trials in the target domain must be known. Zanini \emph{et al.} also proposed improvements to the minimum distance to the Riemannian mean (MDRM) \cite{Barachant2012} classifier and demonstrated the effectiveness of RA and the improved MDRM in both MI and ERP classifications.

Yair \emph{et al.} \cite{Yair2019} proposed a domain adaptation approach using the analytic expression of parallel transport (PT) on the cone manifold of SPD matrices. The goal was to find a common tangent space such that the mappings of $C_t$ and $C_s$ are aligned. It first computes the Riemannian mean $\overline{R}_k$ of the $k$-th domain and then the Riemannian mean $\hat{R}$ of all $\overline{R}_k$. Then, each $\overline{R}_k$ is moved to $\hat{R}$ by PT $\Gamma_{\overline{R}_k\rightarrow\hat{R}}$, and $C_k^n$, the $n$-th covariance matrix in the $k$-th domain, is projected to
\begin{align}
\mathrm{Log}\left(\hat{R}^{-\frac{1}{2}}\Gamma_{\overline{R}_k\rightarrow\hat{R}}\left(C_k^n\right)\hat{R}^{-\frac{1}{2}}\right)
=\mathrm{Log}\left(\overline{R}_k^{-\frac{1}{2}}C_k^n\overline{R}_k^{-\frac{1}{2}}\right).\label{eq:PT}
\end{align}
After the projection step, the covariance matrices in different domains are mapped to the same tangent space, so a classifier built in a source domain can be directly applied to the target domain. Equation (\ref{eq:PT}) is essentially identical to RA in (\ref{eq:RA}), except that (\ref{eq:PT}) works in the tangent space, whereas (\ref{eq:RA}) works in the Riemannian space. Yair \emph{et al.} demonstrated the effectiveness of PT in cross-subject MI classification, sleep stage classification, and mental arithmetic identification.

To make RA more flexible, faster, and completely unsupervised, He and Wu \cite{drwuEA2020} proposed a Euclidean alignment (EA) approach to align EEG trials from different subjects in Euclidean space. Mathematically, for the $k$-th domain, EA computes the reference matrix $\overline{R}_k=\frac{1}{N}\sum_{n=1}^N X_k^n(X_k^n)^\mathrm{T}$, i.e., $\overline{R}_k$ is the arithmetic mean of all covariance matrices in the $k$-th domain (it can also be the Riemannian mean, which is more computationally intensive than the arithmetic mean), then performs the alignment by $\widetilde{X}_k^n=\overline{R}_k^{-\frac{1}{2}}X_k^n$. After EA, the mean covariance matrices of all domains become the identity matrix. Both Euclidean space and Riemannian space feature extraction and classification approaches can then be applied to $\widetilde{X}_k^n$. EA can be viewed as a generalization of Yair \emph{et al.}'s parallel transport approach because the computation of $\overline{R}_k$ in EA is more flexible, and both Euclidean and Riemannian space classifiers can be used after EA. He and Wu demonstrated that EA outperformed RA in both MI and ERP classifications in both offline and simulated online applications.

Rodrigues \emph{et al.} \cite{Rodrigues2019} proposed Riemannian Procrustes analysis (RPA) to accommodate covariant shifts in EEG-based BCIs. It is semi-supervised and requires at least one labelled sample from each target domain class. RPA first matches the statistical distributions of the covariance matrices of the EEG trials from different domains,
{using simple geometrical transformations, namely, translation, scaling, and rotation, in sequence.}
Then, the labelled and transformed data from both domains are concatenated to train a classifier, which is next applied to the transformed and unlabelled target domain samples. Mathematically, it transforms each target domain covariance matrix $C_t^n$ into
\begin{align}
\widetilde{C}_t^n=M_s^{\frac{1}{2}}\left[U^\mathrm{T}
\left(\widetilde{M}_t^{-\frac{1}{2}}C_t^n\widetilde{M}_t^{-\frac{1}{2}}\right)^pU\right]M_s^{\frac{1}{2}}, \label{eq:RPA}
\end{align}
where
\begin{itemize}
\item $\widetilde{M}_t^{-\frac{1}{2}}$ is the geometric mean of the labelled target domain samples, which centres the target domain covariance matrices at the identity matrix.
\item $p=(d/\tilde{d})^{\frac{1}{2}}$ stretches the target domain covariance matrices so that they have the same dispersion as the source domain, in which $d$ and $\tilde{d}$ are the dispersions around the geometric mean of the source domain and the target domain, respectively.
\item $U$ is an orthogonal rotation matrix to be optimized, which minimizes the distance between the class means of the source domain and the translated and stretched target domain.
\item $M_s^{-\frac{1}{2}}$ is the geometric mean of the labelled source domain samples, which ensures that the geometric mean of $\widetilde{C}_t^n$ is the same as that in the source domain.
\end{itemize}
Note that the class label information is only needed in computing $U$. Although $\widetilde{M}_t^{-\frac{1}{2}}$, $p$ and $M_s^{-\frac{1}{2}}$ are also computed from the labeled samples, they do not need the specific class labels.

Clearly, RPA is a generalization of RA. Rodrigues \emph{et al.} \cite{Rodrigues2019} showed that RPA can achieve promising results in cross-subject MI, ERP and SSVEP classification.

Recently, Zhang and Wu \cite{drwuMEKT2020} proposed a manifold embedded knowledge transfer (MEKT) approach, which first aligns the covariance matrices of the EEG trials in the Riemannian manifold, extracts features in the tangent space, and then performs domain adaptation by minimizing the joint probability distribution shift between the source and the target domains while preserving their geometric structures. More specifically, it consists of the following three steps \cite{drwuMEKT2020}:
\begin{enumerate}
\item \emph{Covariance matrix centroid alignment} (CA). Align the centroid of the covariance matrices in each domain, i.e., $\widetilde{C}_s^n=\overline{R}_s^{-\frac{1}{2}} C_s^n \overline{R}_s^{-\frac{1}{2}}$ and $\widetilde{C}_t^n=\overline{R}_t^{-\frac{1}{2}} C_t^n \overline{R}_t^{-f\frac{1}{2}}$, where $\overline{R}_s$ ($\overline{R}_t$) can be the Riemannian mean, the Euclidean mean, or the Log-Euclidean mean of all $C_s^n$ ($C_t^n$). This is essentially a generalization of RA \cite{Zanini2018}. The marginal probability distributions from different domains are brought together after CA.
\item \emph{Tangent space feature extraction}. Map and assemble all $\widetilde{C}_s^n$ ($\widetilde{C}_t^n$) into a tangent space super matrix $\widetilde{X}_s\in\mathbb{R}^{d \times N_s}$ ($\widetilde{X}_t\in\mathbb{R}^{d \times N_t}$), where $d=E(E + 1)/2$ is the dimensionality of the tangent space features.
\item \emph{Mapping matrices identification}. Find the projection matrices $A\in \mathbb{R}^{d \times p}$ and $B\in \mathbb{R}^{d \times p}$, where $p\ll d$ is the dimensionality of the shared subspace, such that $A^\mathrm{T}\widetilde{X}_s$ and $B^\mathrm{T}\widetilde{X}_t$ are similar.
\end{enumerate}
After MEKT, a classifier can be trained on $(A^\mathrm{T}\widetilde{X}_s, \bm{y}_s)$ and applied to $B^\mathrm{T}\widetilde{X}_t$ to estimate their labels.

MEKT can cope with one or more source domains and still be efficient. Zhang and Wu \cite{drwuMEKT2020} also used domain transferability estimation (DTE) to identify the most beneficial source domains, in case there are too many of them. Experiments in cross-subject MI and ERP classification demonstrated that MEKT outperformed several state-of-the-art TL approaches, and DTE can reduce the computational cost to more than half of when the number of source domains is large, with little sacrifice of classification accuracy.

A comparison of the afore-mentioned EEG data alignment approaches, and a new approach \cite{drwuLA2020} introduced later in this section, is given in Table~\ref{tab:alignment}.

\begin{table*}[htbp] \centering  \setlength{\tabcolsep}{1mm}
\caption{Comparison of different EEG data alignment approaches.}   \label{tab:alignment}
\begin{tabular}{c||c|c|c|c|c|c}
\toprule
 & RA \cite{Zanini2018} & PT \cite{Yair2019} & PTA \cite{Rodrigues2019} & EA \cite{drwuEA2020}  & CA \cite{drwuMEKT2020} & LA \cite{drwuLA2020} \\ \midrule

Applicable Paradigm	& MI, ERP &	MI &	MI, ERP, SSVEP &	MI, ERP	& MI, ERP&	MI \\\midrule

Online or Offline &	Both	& Both &	Both & Both &	Both &	Offline\\\midrule

\multicolumn{1}{c||}{\tabincell{c}{Need Labeled Target\\ Domain Trials }} &	\multicolumn{1}{|c|}{\tabincell{c}{No for MI,\\ Yes for ERP}}	& No &	Yes	& No &	No &	Yes\\\midrule

What to Align &	
\multicolumn{1}{|c|}{\tabincell{c}{Riemannian space \\ covariance matrices}} &	\multicolumn{1}{|c|}{\tabincell{c}{Riemannian Tangent \\space features}}	& \multicolumn{1}{|c|}{\tabincell{c}{Riemannian space\\ covariance matrices}}	& \multicolumn{1}{|c|}{\tabincell{c}{Euclidean space\\ EEG trials}} &	\multicolumn{1}{|c|}{\tabincell{c}{Riemannian space\\ covariance matrices}} &	\multicolumn{1}{|c}{\tabincell{c}{Euclidean space\\ EEG trials}}\\\midrule

\multicolumn{1}{c||}{\tabincell{c}{Reference Matrix \\ Calculation}} &	
\multicolumn{1}{|c|}{\tabincell{c}{Riemannian mean\\ of resting state \\ covariance matrices\\ in each domain}}	&
\multicolumn{1}{|c|}{\tabincell{c}{Riemannian mean\\ of all covariance \\matrices in \\each domain}} &	\multicolumn{1}{|c|}{\tabincell{c}{Riemannian mean\\ of all labeled\\ covariance \\matrices in \\each domain}}	&
\multicolumn{1}{|c|}{\tabincell{c}{Euclidean mean \\of all covariance \\matrices in \\each domain}}	& \multicolumn{1}{|c|}{\tabincell{c}{Riemannian,\\ Euclidean, or\\ Log-Euclidean\\ mean of all\\ covariance\\ matrices in\\ each domain}} &	
\multicolumn{1}{|c}{\tabincell{c}{Log-Euclidean\\ mean of labeled \\covariance matrices \\ in each class \\of each domain}}\\\midrule

Classifier	& \multicolumn{1}{|c|}{\tabincell{c}{Riemannian\\ space only}}	& \multicolumn{1}{|c|}{\tabincell{c}{Euclidean\\ space only}}	& \multicolumn{1}{|c|}{\tabincell{c}{Riemannian \\space only}}	& \multicolumn{1}{|c|}{\tabincell{c}{Riemannian or\\ Euclidean space}}	& \multicolumn{1}{|c|}{\tabincell{c}{Riemannian or\\ Euclidean space}}	& \multicolumn{1}{|c}{\tabincell{c}{Riemannian or\\ Euclidean space}}\\\midrule

\multicolumn{1}{c||}{\tabincell{c}{Handle Class\\ Mismatch between\\ Domains}} &	No &	No&	No&	No&	No&	Yes \\\midrule

Computational Cost	&High	&High&	High&	Low&	Low&	Low \\ \bottomrule
\end{tabular}
\end{table*}

Singh \emph{et al.} \cite{Singh2019} proposed a TL approach for estimating the sample covariance matrices, which are used by the MDRM classifier, from a very small number of target domain samples. It first estimates the sample covariance matrix for each class by a weighted average of the sample covariance matrix of the corresponding class from the target domain and that in the source domain. The mixed sample covariance matrix is the sum of the per-class sample covariance matrices. Spatial filters are then computed from the mixed and per-class sample covariance matrices. Next, the covariance matrices of the spatially filtered EEG trials are further filtered by Fisher geodesic discriminant analysis \cite{Barachant2010a} and used as features in the MDRM \cite{Barachant2012} classifier.

Deep learning, which has been very successful in image processing, video analysis, speech recognition and natural language processing, has also started to find applications in EEG-based BCIs. For example, Schirrmeister \emph{et al.} \cite{Schirrmeister2017} proposed two convolutional neural networks (CNNs) for EEG decoding, and showed that both outperformed filter bank common spatial patterns (FBCSP) \cite{Ang2008} in cross-subject MI classification. Lawhern \emph{et al.} \cite{EEGNet} proposed EEGNet, a compact CNN architecture for EEG classification. It can be applied across different BCI paradigms, be trained with very limited data, and generate neurophysiologically interpretable features. EEGNet achieved robust results in both the within-subject and cross-subject classification of MIs and ERPs.

Although the above approaches achieved promising cross-subject classification performance, they did not explicitly use the idea of TL. Currently, a common TL technique for deep learning-based EEG classification \cite{Wang2017e,Wei2019a} is based on fine-tuning with new data from the target session/subject. Unlike concatenating target data with existing source data, the fine-tuning process is established on a pre-trained model and performs iterative learning on a relatively small amount of target data. Although the training data involved is exactly the same as using data concatenation, the prediction performance can be improved significantly.

More specifically, Wu \emph{et al.} \cite{Wu2019a} proposed a parallel multiscale filter CNN for MI classification. It consisted of three layers: a CNN to extract both temporal and spatial features from EEG signals, a feature reduction layer with square and log non-linear functions followed by pooling and dropout, and a dense classification layer fine-tuned on a small amount of calibration data from the target subject. They showed that fine-tuning achieved improved performance in cross-subject transfers.

\subsection{Cross-Device TL}

Xu \emph{et al.} \cite{Xu2020} studied the performance of deep learning in cross-dataset TL. Eight publicly available MI datasets were considered. Although the different datasets used different EEG devices, channels and MI tasks, they only selected three common channels (C3, CZ, C4) and the left-hand and right-hand MI tasks. They applied an online pre-alignment strategy to each EEG trial of each subject by recursively computing the Riemannian mean online and using it as the reference matrix in the EA approach. They showed that online pre-alignment significantly increased the performance of deep learning models in cross-dataset TL.

\subsection{Cross-Task TL}

Both RA and EA assume that the source domains have the same feature space and label space as the target domain, which may not hold in many real-world applications, i.e., they may not be used in cross-task transfers. Recently, He and Wu \cite{drwuLA2020} also proposed a label alignment (LA) approach, which can handle the situation that the source domains have different label spaces from the target domain. For MI-based BCIs, this means the source subjects and the target subject can perform completely different MI tasks (e.g., the source subject may perform left-hand and right-hand MI tasks, whereas the target subject may perform feet and tongue MIs), but the source subjects' data can still be used to facilitate the calibration for a target subject.

When the source and target domain devices are different, LA first selects the source EEG channels that are the most similar to the target EEG channels. Then, it computes the mean covariance matrix of each source domain class and estimates the mean covariance matrix of each target domain class. Next, it re-centres each source domain class at the corresponding estimated class mean of the target domain. Both Euclidean space and Riemannian space feature extraction and classification approaches can next be applied to aligned trials. LA only needs as few as one labelled sample from each target domain class, can be used as a pre-processing step before different feature extraction and classification algorithms, and can be integrated with other TL approaches to achieve an even better performance. He and Wu \cite{drwuLA2020} demonstrated the effectiveness of LA in simultaneous cross-subject, cross-device and cross-task TL in MI classification.

\begin{figure}[htpb]\centering
\includegraphics[width=\linewidth,clip]{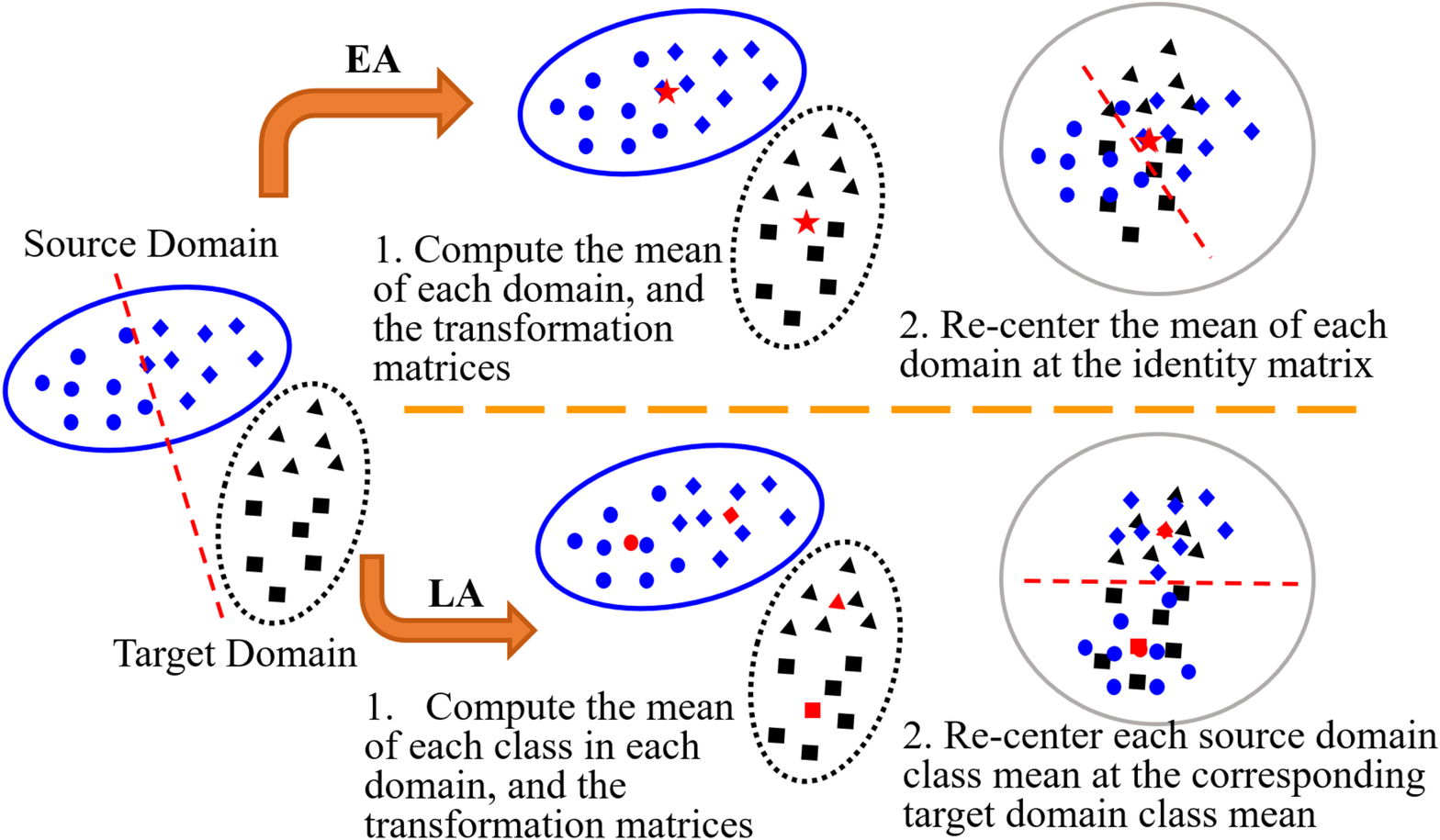}
\caption{Illustration of EA and LA \cite{drwuLA2020}. } \label{fig:LA}
\end{figure}

An illustration of the difference between LA and EA is shown in Fig.~\ref{fig:LA}. To our knowledge, LA is the only cross-task TL work in EEG-based BCIs and the most complicated TL scenario (simultaneous cross-subject, cross-device and cross-task TL) considered in the literature so far.

\section{TL in ERP-based BCIs} \label{ect:ERP}

This section reviews recent TL approaches in ERP-based BCIs. Many approaches introduced in the previous section, e.g., RA, EA, RPA and EEGNet, can also be used here. To avoid duplication, we only include approaches not introduced in the previous section here. Because there were no publications on cross-task TL in ERP-based BCIs, we do not have a ``Cross-Task TL" subsection.

\subsection{Cross-Subject/Session TL} \label{sect:wAR}

Waytowich \emph{et al.} \cite{Waytowich2016} proposed unsupervised spectral transfer method using information geometry (STIG) for subject-independent ERP-based BCIs. STIG uses a spectral meta-learner \cite{Parisi2014} to combine predictions from an ensemble of MDRM classifiers on data from individual source subjects. Experiments on single-trial ERP classification demonstrated that STIG significantly outperformed some calibration-free approaches and traditional within-subject calibration approaches when limited data were available in both offline and online ERP classifications.

Wu \cite{drwuTHMS2017} proposed weighted adaptation regularization (wAR) for cross-subject transfers in ERP-based BCIs in both online and offline settings. Mathematically, wAR learns the following classifier directly:
\begin{align}
\argmin\limits_f&\sum_{n=1}^{N_s}w_s^n\ell(f(X_s^n),y_s^n)+w_t\sum_{n=1}^{N_l}w_t^n\ell(f(X_t^n),y_t^n) \nonumber\\
& +\sigma\|f\|_K^2+\lambda_P D_{f,K}(P_s(X_s),P_t(X_t))\nonumber \\
&+\lambda_Q D_{f,K}(P_s(X_s|y_s),P_t(X_t|y_t)) \label{eq:wAR}
\end{align}
where $\ell$ is a loss function, $w_t$ is the overall weight of target domain samples, $K$ is a kernel function, and $\sigma$, $\lambda_P$ and $\lambda_Q$ are non-negative regularization parameters. $w_s^n$ and $w_t^n$ are the weights for the $n$-th sample in the source domain and the target domain, respectively, to balance the number of positive and negative samples in the corresponding domain.

Briefly, the five terms in (\ref{eq:wAR}) minimize the fitting loss in the source domain, the fitting loss in the target domain, the structural risk of the classifier, the distance between the marginal probability distributions $P_s(X_s)$ and $P_t(X_t)$, and the distance between the conditional probability distributions $P_s(X_s|y_s)$ and $P_t(X_t|y_t)$. Experiments on single-trial visual evoked potential classification demonstrated that both online and offline wAR algorithms were effective. Wu \cite{drwuTHMS2017} also proposed a source domain selection approach, which selects the most beneficial source subjects for transferring. It can reduce the computational cost of wAR by $\sim$50\% without sacrificing the classification performance.

Qi \emph{et al.} \cite{Qi2018} performed cross-subject TL on a P300 speller to reduce the calibration time. A small set of ERP epochs from the target subject was used as a reference to compute the Riemannian distance to each source ERP sample from an existing data pool. The most similar ones were selected to train a classifier and were applied to the target subject.

Jin \emph{et al.} \cite{Jin2020} used a generic model set to reduce the calibration time in P300-based BCIs. Filtered EEG data from 116 participants were assembled into a data matrix, principal component analysis (PCA) was used to reduce the dimensionality of the time domain features, and then the 116 participants were clustered into 10 groups by $k$-means clustering. A weighted linear discriminant analysis (WLDA) classifier was then trained for each cluster. These 10 classifiers formed the generic model set. For a new subject, a few calibration samples were acquired, and an online linear discriminant (OLDA) model was trained. The OLDA model was matched to the closest WLDA model, which was then selected as the model for the new subject.

Deep learning has also been used in ERP classification. Inspired by generative adversarial networks (GANs) \cite{Goodfellow2014}, Ming \emph{et al.} \cite{drwuASC2019} proposed a subject adaptation network (SAN) to mitigate individual differences in EEGs. Based on the characteristics of the application, they designed an artificial low-dimensional distribution and forced the transformed EEG features to approximate it. For example, for two-class visual evoked potential classification, the artificial distribution is bimodal, and the area of each modal is proportional to the number of samples in the corresponding class. Experiments on cross-subject visual evoked potential classification demonstrated that SAN outperformed a support vector machine (SVM) and EEGNet.

\subsection{Cross-Device TL}

Wu \emph{et al.} \cite{drwuTNSRE2016} proposed active weighted adaptation regularization (AwAR) for cross-device TL. It integrates wAR (introduced in Section~\ref{sect:wAR}), which uses labelled data from the previous device and handles class imbalance, and active learning \cite{Settles2009}, which selects the most informative samples from the new device to label. Only the common channels were used in wAR, but all the channels of the new device can be used in active learning to achieve better performance. Experiments on single-trial visual evoked potential classification using three different EEG devices with different numbers of electrodes showed that AwAR can significantly reduce the calibration data requirement for a new device in offline calibration.

To our knowledge, this is the only study on cross-device TL in ERP-based BCIs.

\section{TL in SSVEP-based BCIs} \label{sect:SSVEP}

This section reviews recent TL approaches in SSVEP-based BCIs. Because there were no publications on cross-task TL in SSVEP-based BCIs, we do not have a ``Cross-Task TL" subsection. Overall, many fewer TL studies on SSVEPs have been performed compared with MI tasks and ERPs.

\subsection{Cross-Subject/Session TL}

Waytowich \emph{et al.} \cite{Waytowich2018} proposed Compact-CNN, which is essentially the EEGNet \cite{EEGNet} approach introduced in Section~\ref{sect:CSSMI}, for 12-class SSVEP classification without the need for any user-specific calibration. It outperformed state-of-the-art hand-crafted approaches using canonical correlation analysis (CCA) and Combined-CCA.

Rodrigues \emph{et al.} \cite{Rodrigues2019} proposed RPA, which can also be used in cross-subject transfer of SSVEP-based BCIs. Since it has been introduced in Section~\ref{sect:CSSMI}, it is not repeated here.

\subsection{Cross-Device TL}

Nakanishi \emph{et al.} \cite{Nakanishi2020} proposed a cross-device TL algorithm for reducing the calibration effort in an SSVEP-based BCI speller. It first computes a set of spatial filters by channel averaging, CCA, or task-related component analysis and then concatenates them to form a filter matrix $W$. The average trial of Class~$c$ of the source domain is computed and filtered by $W$ to obtain $Z_c$. Let $X_t$ be a single trial to be classified in the target domain. Its spatial filter matrix $W_c$ is then computed by
\begin{align}
W_c=\argmin_W\left\|Z_c-W^\mathrm{T}X_t\right\|_2^2;
\end{align}
i.e., $W_c=(X_tX_t^\mathrm{T})^{-1}X_tZ_c^\mathrm{T}$. Then, Pearson's correlation coefficients between $W_c^\mathrm{T}X_t$ and $Z_c$ are computed as $r_c^{(1)}$, and canonical correlation coefficients between $X_t$ and computer-generated SSVEP models $Y_c$ are computed as $r_c^{(2)}$. The two feature values are combined as
\begin{align}
\rho_c=\sum_{i=1}^2 \mathrm{sign}\left(r_c^{(i)}\right)\cdot \left(r_c^{(i)}\right)^2,
\end{align}
and the target class is identified as $\argmax\limits_c \rho_c$.

To our knowledge, this is the only study on cross-device TL in SSVEP-based BCIs.

\section{TL in aBCIs} \label{sect:aBCI}

Recently, there has been a fast-growing research interest in aBCIs \cite{Muhl2014,AlNafjan2017,Shen2019,Alarcao2019}. Emotions can be represented by discrete categories \cite{Ekman1971} (e.g., happy, sad, and angry) and by continuous values in the 2D space of arousal and valence \cite{Russell1980} or the 3D space of arousal, valence, and dominance \cite{Mehrabian1980}. Therefore, there can be both classification and regression problems in aBCIs. However, the current literature focused exclusively on classification problems.

Most studies used the publicly available
{DEAP}
\cite{Koelstra2012} and SEED \cite{Zheng2015} datasets. DEAP consists of 32-channel EEGs recorded by a BioSemi ActiveTwo device from 32 subjects while they were watching minute-long music videos, whereas SEED consists of 62-channel EEGs recorded by an ESI NeuroScan device from 15 subjects while they were watching 4-minute movie clips. By using SEED, Zheng \emph{et al.} \cite{Zheng2019} investigated whether stable EEG patterns exist over time for emotion recognition. Using differential entropy features, they found that stable patterns did exhibit consistency across sessions and subjects. Thus, it is possible to perform TL in aBCIs.

This section reviews the latest progress on TL in aBCIs. Because there were no publications on cross-task TL in aBCIs, we do not have a ``Cross-Task TL'' subsection.

\subsection{Cross-Subject/Session TL}


Chai \emph{et al.} \cite{Chai2017} proposed adaptive subspace feature matching (ASFM) for cross-subject and cross-session transfer in offline and simulated online EEG-based emotion classification. Differential entropy features were used. ASFM first performs PCA of the source domain and the target domain separately. Let $Z_s$ ($Z_t$) be the $d$ leading principal components in the source (target) domain, which form the corresponding subspace. Then, ASFM transforms the source domain subspace to $Z_sZ_s^\mathrm{T}Z_t$ and projects the source data into it. The target data are projected directly into $Z_t$. In this way, the marginal distribution discrepancy between the two domains is reduced. Next, an iterative pseudo-label refinement strategy is used to train a logistic regression classifier using the labelled source domain samples and pseudo-labelled target domain samples, which can be directly applied to unlabelled target domain samples.

Lin and Jung \cite{Lin2017a} proposed a conditional TL (cTL) framework to facilitate positive cross-subject transfers in aBCIs. Five differential laterality features (topoplots), corresponding to five different frequency bands, from each EEG channel are extracted. The cTL method first computes the classification accuracy by using the target subject's data only and performs transfer only if that accuracy is below the chance level. Then, it uses ReliefF \cite{Kononenko1994} to select a few of the most emotion-relevant features in the target domain and calculates their correlations with the corresponding features in each source domain to select a few of the most similar (correlated) source subjects. Next, the target domain data and the selected source domain data are concatenated to train a classifier.

Lin \emph{et al.} \cite{Lin2017} proposed a robust PCA (RPCA)-based \cite{Candes2011} signal filtering strategy and validated its performance in cross-day binary emotion classification. RPCA decomposes an input matrix into the sum of a low-rank matrix and a sparse matrix. The former accounts for the relatively regular profiles of the input signals, whereas the latter accounts for its deviant events.
{Lin \emph{et al.} showed that the RPCA-decomposed sparse signals filtered from the background EEG activity contributed more to the inter-day variability and that the predominately captured the EEG oscillations of emotional responses behaved relatively consistently across days.}

Li \emph{et al.} \cite{Li2018a} extracted nine types of time-frequency domain features (the peak-to-peak mean, mean square, variance, Hjorth activity, Hjorth mobility, Hjorth complexity, maximum power spectral frequency, maximum power spectral density, power sum) and nine types of non-	linear dynamical system features (the approximate entropy, C0 complexity, correlation dimension, Kolmogorov entropy, Lyapunov exponent, permutation entropy, singular entropy, Shannon entropy, spectral entropy) from EEG measurements. Through automatic and manual feature selection, they verified the effectiveness and performance of the upper bounds of those features in cross-subject emotion classification on DEAP and SEED. They found that L1-norm penalty-based feature selection achieved robust performance on both datasets, and the Hjorth mobility in the beta rhythm achieved the highest mean classification accuracy.

Liu \emph{et al.} \cite{Liu2018a} performed cross-day EEG-based emotion classification. Seventeen subjects watched 6-9 emotional movie clips on five different days over one month. Spectral powers of the delta, theta, alpha, beta, low and high gamma bands were computed for each of the 60 channels as initial features, and then recursive feature elimination was used for feature selection. In cross-day classification, the data from a subset of the five days were used by an SVM to classify the data from the remaining days. They showed that EEG variability could impair the emotion classification performance dramatically, and using data from more days during training could significantly improve the generalization performance.

Yang \emph{et al.} \cite{Yang2019} studied cross-subject emotion classification on DEAP and SEED. Ten linear and non-linear features (the Hjorth activity, Hjorth mobility, Hjorth complexity, standard deviation, PSD alpha, PSD beta, PSD gamma, PSD theta, sample entropy, and wavelet entropy) were extracted from each channel and concatenated. Then, sequential backward feature selection and significance test were used for feature selection, and an RBF SVM was used as the classifier.

Li \emph{et al.} \cite{Li2020a} considered cross-subject EEG emotion classification for both supervised (the target subject has some labelled samples) and semi-supervised (the target subject has some labelled samples and unlabelled samples) scenarios. We briefly introduce only their best-performing supervised approach here. Multiple source domains are assumed. They first performed source selection by training a classifier in each source domain and compute its classification accuracy on the labelled samples in the target domain. These accuracies were then sorted to select the top few source subjects. A style transfer mapping was learned between the target and each selected source. For each selected source subject, they performed SVM classification on his/her data, removed the support vectors (because they are near the decision boundary and hence uncertain), performed $k$-means clustering on the remaining samples to obtain the prototypes, and mapped each target domain labelled sample feature vector $\bm{x}_t^n$ to the nearest prototype in the same class by the following mapping:
\begin{align}
\min\limits_{A,b}\sum_{i=1}^n \|A\bm{x}_t^n+b-d^n\|_2^2+\beta\|A-I\|_F^2+\gamma\|b\|_2^2,
\end{align}
where $\bm{d}_n$ is the nearest prototype in the same class of the source domain, and $\beta$ and $\gamma$ are hyperparameters. A new, unlabelled sample in the target domain is first mapped to each selected source domain and then classified by a classifier trained in the corresponding source domain. The classification results from all source domains were then weighted averaged, where the weights were determined by the accuracies of the source domain classifiers.

Deep learning has also been gaining popularity in aBCI.

Chai \emph{et al.} \cite{Chai2016} proposed a subspace alignment autoencoder (SAAE) for cross-subject and cross-session transfer in EEG-based emotion classification. First, differential entropy features from both domains were transformed into a domain-invariant subspace using a stacked autoencoder. Then, kernel PCA, graph regularization and maximum mean discrepancy were used to reduce the feature distribution discrepancy between the two domains. After that, a classifier trained in the source domain can be directly applied to the target domain.

Yin and Zhang \cite{Yin2017a} proposed an adaptive stacked denoising autoencoder (SDAE) for cross-session binary classification of mental workload levels from EEG. The weights of the shallow hidden neurons of the SDAE were adaptively updated during the testing phase using augmented testing samples and their pseudo-labels.

Zheng \emph{et al.} \cite{Zheng2019a} presented EmotionMeter, a multi-modal emotion recognition framework that combines brain waves and eye movements to classify four emotions (fear, sadness, happiness, and neutrality). They adopted a bimodal deep autoencoder to extract the shared representations of both EEGs and eye movements. Experimental results demonstrated that modality fusion combining EEG and eye movements with multi-modal deep learning can significantly enhance emotion recognition accuracy compared with a single modality. They also investigated the complementary characteristics of EEGs and eye movements for emotion recognition and the stability of EmotionMeter across sessions. They found that EEGs and eye movements have important complementary characteristics, e.g., EEGs have the advantage of classifying happy emotion (80\%) compared with eye movements (67\%), whereas eye movements outperform EEGs in recognizing fear emotion (67\% versus 65\%).

Fahimi \emph{et al.} \cite{Fahimi2019} performed cross-subject attention classification. They first trained a CNN by combining EEG data from the source subjects and then fine-tuned it by using some calibration data from the target subject. The inputs were raw EEGs, bandpass filtered EEGs, and decomposed EEGs (delta, theta, alpha, beta and gamma bands).

Li \emph{et al.} \cite{Li2020b} proposed R2G-STNN, which consists of spatial and temporal neural networks with regional-to-global hierarchical feature learning, to learn discriminative spatial-temporal EEG features for subject-independent emotion classification. To learn the spatial features, a bidirectional long short-term memory (LSTM) network was used to capture the intrinsic spatial relationships of EEG electrodes within and between different brain regions. A region-attention layer was also introduced to learn the weights of different brain regions. A domain discriminator working corporately with the classifier was used to reduce domain shift between training and testing.

Li \emph{et al.} \cite{Li2020c} further proposed an improved bi-hemisphere domain adversarial neural network
(BiDANN-S) for subject-independent emotion classification. Inspired by the neuroscience findings that the left and right hemispheres of the human brain are asymmetric to the emotional response, BiDANN-S uses a global and
two local domain discriminators working adversarially with a classifier to learn discriminative emotional features for each hemisphere. To improve the generalization performance and to facilitate subject-independent EEG emotion classification, it also tries to reduce the possible domain differences in each hemisphere between the source and target domains and ensure that the extracted EEG features are robust to subject variations.

Li \emph{et al.} \cite{Li2020d} proposed a neural network model for cross-subject/session EEG emotion recognition, which does not require label information in the target domain. The neural network was optimized by minimizing the classification error in the source domain while making the source and target domains similar in their latent representations. Adversarial training was used to adapt the marginal distributions in the early layers, and association reinforcement was performed to adapt the conditional distributions in the last few layers. In this way, it achieved joint distribution adaptation \cite{Long2013}.

Song \emph{et al.} \cite{Song2020} proposed a dynamical graph convolutional neural network (DGCNN) for subject-dependent and subject-independent emotion classification. Each EEG channel was represented as a node in the DGCNN, and differential entropy features from five frequency bands were used as inputs. After graph filtering, a $1\times1$ convolution layer learned the discriminative features among the five frequency bands. A ReLU activation function was adopted to ensure that the outputs of the graph filtering layer are non-negative. The outputs of the activation function were sent to a multilayer dense network for classification.

Appriou \emph{et al.} \cite{Appriou2020} compared several modern machine learning algorithms on subject-specific and subject-independent cognitive and emotional state classification, including Riemannian approaches and a CNN. They found that the CNN performed the best in both subject-specific and subject-independent workload classification. A filter bank tangent space classifier (FBTSC) was also proposed. It first filters an EEG into several different frequency bands. For each band, it computes the covariance matrices of the EEG trials, projects them onto the tangent space at their mean, and then applies a Euclidean space classifier. FBTSC achieved the best performance in subject-specific emotion (valance and arousal) classification.

\subsection{Cross-Device TL}

Lan \emph{et al.} \cite{Lan2019} considered cross-dataset transfers between DEAP and SEED, which have different numbers of subjects, and were recorded using different EEG devices with different numbers of electrodes. They used only three trials (one positive, one neutral, and one negative) from 14 selected subjects in DEAP and only the 32 common channels between the two datasets. Five differential entropy features in five different frequency bands (delta, theta, alpha, beta, and gamma) were extracted from each channel and concatenated as features. Experiments showed that domain adaptation, particularly transfer component analysis \cite{Pan2010a} and maximum independence domain adaptation \cite{Yan2018}, can effectively improve the classification accuracies compared to the baseline.

Lin \cite{Lin2020} proposed RPCA-embedded TL to generate a personalized cross-day emotion classifier with less labelled data while obviating intra- and inter-individual differences. The source dataset consists of 12 subjects using a 14-channel Emotiv EPOC device, and the target dataset consists of 26 different subjects using a 30-channel Neuroscan Quik-Cap. Twelve of the 26 channels of Quik-Cap were first selected to align with 12 of the 14 selected channels from the EPOC device. The Quik-Cap EEG signals were also down-sampled and filtered to match those of the EPOC device. Five frequency band (delta, theta, alpha, beta, gamma) features from each of the six left-right channel pairs (e.g., AF3-AF4, F7-F8), four fronto-posterior pairs (e.g., AF3-O1, F7-P7) and the 12 selected channels were extracted, resulting in a 120D feature vector for each trial. Similar to \cite{Lin2017}, the sparse RPCA matrix of the feature matrix was used as the final feature. The Riemannian distance between the trials of each source subject and the target subject was computed as a dissimilarity measure to select the most similar source subjects, whose trials were combined with the trials from the target subject to train an SVM classifier.

Zheng \emph{et al.} \cite{Zheng2020a} considered an interesting cross-device (or cross-modality) and cross-subject TL scenario in which the target subject's eye tracking data were used to enhance the performance of cross-subject EEG-based affect classification. It is a 3-step procedure. First, multiple individual emotion classifiers are trained for the source subjects. Second, a regression function is learned to model the relationship between the data distribution and classifier parameters. Third, a target classifier is constructed using the target feature distribution and the distribution-to-classifier mapping. This heterogeneous TL approach achieved comparable performance with homogeneous EEG-based models and scanpath-based models. To our knowledge, this is the first study that transfers between two different types of signals.

Deep learning has also been used in cross-device TL in aBCIs. EEG trials are usually transformed to some sort of images before input to the deep learning model. In this way, EEG signals from different devices can be made consistent.

Siddharth \emph{et al.} \cite{Siddharth2020} performed multimodality (e.g., EEG, ECG, face, etc.) cross-dataset emotion classification, e.g., training on DEAP and testing on the MAHNOB-HCI database \cite{Soleymani2012}. We only briefly introduce their EEG-based deep learning approach here, which works for datasets with different numbers and placements of electrodes, different sampling rates, etc. The EEG power spectral densities (PSDs) in the theta, alpha and beta bands were used to plot three topographies for each trial. Then, each topography was considered a component of a colour image and weighted by the ratio of alpha blending to form the colour image. In this way, one colour image representing the topographic PSD was obtained for each trial, and the images obtained from different EEG devices can be directly combined or compared. A pre-trained VGG-16 network was used to extract 4,096 features from each image, whose number was later reduced to 30 by PCA. An extreme learning machine was used as the classifier for final classification.

Cimtay and Ekmekcioglu \cite{Cimtay2020} used a pre-trained state-of-the-art CNN model, Inception-ResNet-v2, for cross-subject and cross-dataset transfers. Since Inception-ResNet-v2 requires the input data size to be $(N_1,N,3)$, where $N_1\ge 75$ is the number of EEG channels and $N\ge 75$ is the number of time domain samples, when the number of EEG channels is be less than 75, they increased the number of channels by adding noisy copies of them (Gaussian random noise was used) to reach $N_1=80$. This process was repeated three times so that each trial became a $80\times300\times3$ matrix, which was then used as the input to Inception-ResNet-v2. They also added a global average pooling layer and five dense layers after Inception-ResNet-v2 for classification.

\section{TL in BCI Regression Problems} \label{sect:Regression}

There are many important BCI regression problems, e.g., driver drowsiness estimation \cite{drwuFWET2019,drwuTITS2020,drwuTFS2017}, vigilance estimation \cite{Shi2013,Zheng2017a,Zheng2020}, and user reaction time estimation \cite{drwuRG2017}, which were not adequately addressed in previous reviews. This section fills this gap. Because there were no publications on cross-device and cross-task TL in BCI regression problems, we do not have subsections on them.

\subsection{Cross-Subject/Session TL}

Wu \emph{et al.} \cite{drwuTFS2017} proposed a novel online weighted adaptation regularization for regression (OwARR) algorithm to reduce the amount of subject-specific calibration data in EEG-based driver drowsiness estimation and a source domain selection approach to save approximately half of its computational cost. OwARR minimizes the following loss function, similar to wAR \cite{drwuTNSRE2016}:
\begin{align}
\min\limits_f&\sum_{n=1}^{N_s}\ \left(y_s^n-f(X_s^n)\right)^2+w_t\sum_{n=1}^{N_l}\left(y_t^n-f(X_t^n)\right)^2 \nonumber \\
&+\lambda \left[d(P_s(X_s),P_t(X_t))+ d(P_s(X_s|y_s),P_t(X_t|y_t))\right]\nonumber\\
&- \gamma \tilde{r}^2(y,f(X)) \label{eq:OwARR}
\end{align}
where $\lambda$ and $\gamma$ are non-negative regularization parameters and $w_t$ is the overall weight for target domain samples. $\tilde{r}^2(y,f(X))$ approximates the sample Pearson correlation coefficient between $y$ and $f(X)$. Fuzzy sets were used to define fuzzy classes so that $d(P_s(X_s|y_s),P_t(X_t|y_t))$ can be efficiently computed. The five terms in (\ref{eq:OwARR}) minimize the fitting error in the source domain, the fitting error in the target domain, the distance between the marginal probability distributions, the distance between the conditional probability distributions, and the estimated sample Pearson correlation coefficient between $y$ and $f(X)$. Wu \emph{et al.} \cite{drwuTFS2017} showed that OwARR and OwARR with source domain selection can achieve significantly smaller estimation errors than several other cross-subject TL approaches.

Jiang \emph{et al.} \cite{drwuTITS2020} further extended OwARR to multi-view learning, where the first view included theta band powers from all channels, and the second view converted the first view into dBs and removed some bad channels. A TSK fuzzy system was used as the regression model, optimized by minimizing (\ref{eq:OwARR}) for both views simultaneously and adding an additional term to enforce the consistency between the two views (the estimation from one view should be close to that from the other view). They demonstrated that the proposed approach outperformed a domain adaptation with a model fusion approach \cite{drwuaBCI2015} in cross-subject TL.

Wei \emph{et al.} \cite{Wei2018} also performed cross-subject driver drowsiness estimation. Their procedure consisted of three steps: 1) Ranking. For each source subject, it computed six distance measures (Euclidean distance, correlation distance, Chebyshev distance, cosine distance, Kullback-Leibler divergence, and transferability-based distance) between his/her own alert baseline (the first 10 trials) power distribution and all other source subjects' distributions and the cross-subject model performance (XP), which is the transferability of other source subjects on the current subject. A support vector regression (SVR) model was then trained to predict XP from the distance measures. In this way, given a target subject with a few calibration trials, the XP of the source subjects can be computed and ranked. 2) Fusion: a weighted average was used to combine the source models, where the weights were determined from a modified logistic function optimized on the source subjects. 3) Re-calibration: the weighted average was subtracted by an offset, estimated as the median of the initial 10 calibration trials (i.e., the alert baseline) from the target subject. They showed that this approach can result in a 90\% calibration time reduction in driver drowsiness index estimation.

Chen \emph{et al.} \cite{Chen2019} integrated feature selection and an adaptation regularization-based TL (ARTL) \cite{Long2014} classifier for cross-subject driver status classification. The most novel part is feature selection, which extends the traditional ReliefF \cite{Kononenko1994} and minimum redundancy maximum relevancy (mRMR) \cite{Peng2005} to class separation and domain fusion (CSDF)-ReliefF and CSDF-mRMR, which consider both the class separability and the domain similarity, i.e., the selected feature subset should simultaneously maximize the distinction among different classes and minimize the difference among different domains. The ranks of the features from different feature selection algorithms were then fused to identify the best feature set, which was used in ARTL for classification.

Deep learning has also been used in BCI regression problems.

Ming \emph{et al.} \cite{drwuEEG2018} proposed a stacked differentiable neural computer and demonstrated its effectiveness in cross-subject EEG-based mind load estimation and reaction time estimation. The original long short-term memory network controller in differentiable neural computers was replaced by a recurrent convolutional network controller, and the memory-accessing structures were also adjusted for processing EEG topographic data.

Cui \emph{et al.} \cite{drwuFWET2019} proposed a subject-independent TL approach, feature weighted episodic training (FWET), to completely eliminate the calibration requirement in cross-subject transfers in EEG-based driver drowsiness estimation. It integrates feature weighting to learn the importance of different features and episodic training for domain generalization. Episodic training considers the conditional distributions $P(y_s|f(X_s))$ directly and trains a regression network $f$ that aligns $P(y_s|f(X_s))$ in all the source domains, which usually generalizes well to the unseen target domain $\mathcal{D}_t$. It first establishes a subject-specific feature transformation model $f_{\bm{\theta}_{s}}$ and a subject-specific regression model $f_{\bm{\psi}_s}$ for each source subject to learn the domain-specific information, then trains a feature transformation model $f_{\bm{\theta}}$ that makes the transformed features from Subject~$s$ still perform well when applied to a regressor $f_{\bm{\psi}_j}$ trained on Subject~$j$ ($j\neq s$). The overall loss function of episodic training, when Subject $s$'s data are fed into Subject~$j$'s regressor, is:
\begin{align}
\ell_{s,j} = &\sum_{n=1}^{N_s}\ell(y_s^n,f_{\bm{\psi}}(f_{\bm{\theta}}(X_s^n)))\nonumber\\
&+\lambda \sum_{n=1}^{N_s}\ell(y_s^n, \overline{f}_{\bm{\psi}_j}(f_{\bm{\theta}}(X_s^n))), \label{eq:FWET}
\end{align}
where $\overline{f}_{\bm{\psi}_j}$ means that $f_{\bm{\psi}_j}$ is not updated during backpropagation. Once the optimal $f_{\bm{\psi}}$ and $f_{\bm{\theta}}$ are obtained, the prediction for $X_t$ is $\hat{y}_t=f_{\bm{\psi}}(f_{\bm{\theta}}(X_t))$.

\section{TL in Adversarial Attacks of EEG-Based BCIs} \label{sect:Attack}

Adversarial attacks of EEG-based BCIs represent one of the latest developments in BCIs. It was first studied by Zhang and Wu \cite{drwuBCIAttack2019}. They found that adversarial perturbations, which are deliberately designed tiny perturbations, can be added to normal EEG trials to fool the machine learning model and cause dramatic performance degradation. Both traditional machine learning models and deep learning models, as well as both classifiers and regression models in EEG-based BCIs, can be attacked.

\begin{figure}[htbp]   \centering
\includegraphics[width=.7\linewidth,clip]{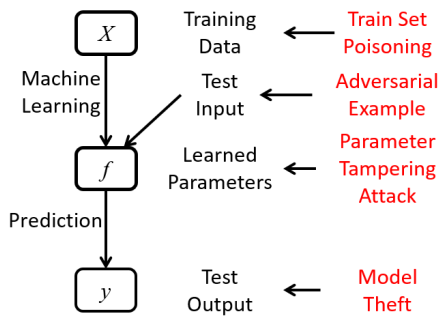}
\caption{Attack strategies to different components of a machine learning model.} \label{fig:AA}
\end{figure}

Adversarial attacks can target different components of a machine learning model, e.g., training data, model parameters, test data, and test output, as shown in Fig.~\ref{fig:AA}. To date, only adversarial examples (benign examples contaminated by adversarial perturbations) targeting the test inputs have been investigated in EEG-based BCIs, so this section only considers adversarial examples.

\begin{figure}[htbp]   \centering
\includegraphics[width=.7\linewidth,clip]{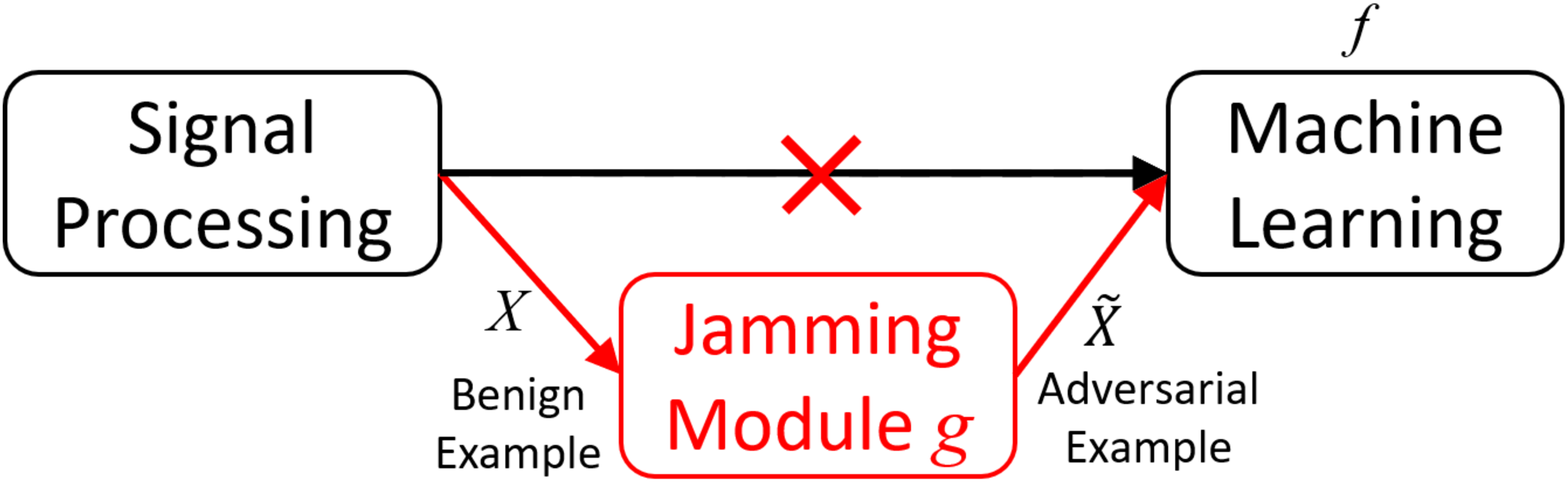}
\caption{Adversarial example generation scheme~ \cite{drwuBCIAttack2019}.} \label{fig:AA2}
\end{figure}

A more detailed illustration of the adversarial example attack scheme is shown in Fig.~\ref{fig:AA2}. A jamming module is injected between signal processing and machine learning to generate adversarial examples.

Table~\ref{tab:summary} shows the three attack types in EEG-based BCIs. White-box attacks know all information about the victim model, including its architecture and parameters, and hence are the easiest to perform. Black-box attacks know nothing about the victim model but can only supply inputs to it and observe its output and hence are the most challenging to perform.

\begin{table}[htbp] \center
\caption{Summary of the three attack types in EEG-based BCIs~\cite{drwuBCIAttack2019}.}   \label{tab:summary}
\begin{tabular}{c|ccc}
\toprule
Victim Model & White-Box & Grey-Box  & Black-Box  \\
Information & Attacks & Attacks & Attacks \\ \midrule
Know its architecture & $\checkmark$ & $\times$ & $\times$ \\
Know its parameters $\boldsymbol{\theta}$ & $\checkmark$ & $\times$ & $\times$ \\
Know its training data &  $-$ & $\checkmark$ & $\times$ \\
Can observe its response &  $-$ & $-$ & $\checkmark$ \\ \bottomrule
\end{tabular}
\end{table}

\subsection{Cross-Model Attacks}

Different from the cross-subject/session/device/task TL scenarios considered in the previous five sections, adversarial attacks in BCIs so far mainly considered cross-model attacks\footnote{Existing publications \cite{drwuBCIAttack2019,drwuTAR2019} also considered \emph{cross-subject} attacks, but the meaning of cross-subject in adversarial attacks is different from the cross-subject TL setting in previous sections: in adversarial attacks, cross-subject means that the same machine learning model is used by all subjects, but the scheme for generating adversarial examples is designed on some subjects and applied to another subject. It assumes that the victim machine learning model works well for all subjects.}, where adversarial examples generated from one machine learning model are used to attack another model. This assumption is necessary in grey-box and black-box attacks because the victim model is unknown, and the attacker needs to construct its own model (called the substitute model) to approximate the victim model.

Interestingly, cross-model attacks can be performed without explicitly considering TL. They are usually achieved by making use of the transferability of adversarial examples \cite{Szegedy2014}, i.e., adversarial examples generated by one machine learning model may also be used to fool a different model. The fundamental reason behind this property is still unclear, but it does not hinder people from making use of it.

For example, Zhang and Wu \cite{drwuBCIAttack2019} proposed unsupervised fast gradient sign methods, which can effectively perform white-box, grey-box and black-box attacks on deep learning classifiers. Two BCI paradigms, i.e., MI and ERP, and three popular deep learning models, i.e., EEGNet, Deep ConvNet and Shallow ConvNet, were considered. Meng \emph{et al.} \cite{drwuTAR2019} further showed that the transferability of adversarial examples can also be used to attack regression models in BCIs; e.g., adversarial examples designed from a multi-layer perceptron neural network can be used to attack a ridge regression model, and vice versa, in EEG-based user reaction time estimation.

\section{Conclusions} \label{sect:Conclusions}

This paper has reviewed recently proposed TL approaches in EEG-based BCIs, according to six different paradigms and applications: MI, ERP, SSVEP, aBCI, regression problems, and adversarial attacks. TL algorithms are grouped into cross-subject/session, cross-device and cross-task approaches and introduced separately. Connections among similar approaches are also pointed out.

The following observations and conclusions can be made, which may point to some future research directions:
\begin{enumerate}
\item Among the three classic BCI paradigms (MI, ERP and SSVEP), SSVEP seems to receive the least amount of attention. Very few TL approaches have been proposed recently. One reason may be that MI and ERP are very similar, so many TL approaches developed for MI can be applied to ERPs directly or with little modification, e.g., RA, EA, RPA and EEGNet, whereas SSVEP is a quite different paradigm.

\item Two new applications of EEG-based BCIs, i.e., aBCI and regression problems, have been attracting increasing research interest. Interestingly, both of them are passive BCIs \cite{Arico2018}. Although both classification and regression problems can be formulated in aBCIs, existing research has focused almost exclusively on classification problems.

\item Adversarial attacks, one of the latest developments in EEG-based BCIs, can be performed across different machine learning models by utilizing the transferability of adversarial examples. However, explicitly considering TL between different domains may further improve the attack performance. For example, in black-box attacks, TL can make use of publicly available datasets to reduce the number of queries to the victim model or, in other words, to better approximate the victim model given the same number of queries.

\item Most TL studies focused on cross-subject/session transfers. Cross-device transfers have started to attract attention, but cross-task transfers remain largely unexplored. To our knowledge, there has been only one such study \cite{drwuLA2020} since 2016. Effective cross-device and cross-task transfers would make EEG-based BCIs much more practical.

\item Among various TL approaches, Riemannian geometry and deep learning are emerging and gaining momentum, each of which has a group of approaches proposed.

\item Although most research on TL in BCIs has focused on classifiers or regression models, i.e., at the pattern recognition stage, TL in BCIs can also be performed in trial alignment, e.g., RA, EA, LA and RPA, in signal filtering, e.g., transfer kernel common spatial patterns \cite{Dai2018}, and in feature extraction/selection, e.g., CSDF-ReliefF and CSDF-mRMR \cite{Chen2019}. Additionally, these TL-based individual components can also be assembled into a complete machine learning pipeline to achieve even better performance. For example, EA and LA data alignment schemes have been combined with TL classifiers \cite{drwuEA2020,drwuLA2020}, and CSDF-ReliefF and CSDF-mRMR feature selection approaches have also been integrated with TL classifiers \cite{Chen2019}.

\item TL can also be integrated with other machine learning approaches, e.g., active learning \cite{Settles2009}, for improved performance \cite{drwuPLOS2013,drwuTNSRE2016}.
\end{enumerate}



\end{document}